\documentclass[journal=ancham,manuscript=article]{achemso}

\usepackage[version=3]{mhchem} 
\usepackage[graphicx]{}
\usepackage[textcomp]{}

\newcommand*{\registered}{\textsuperscript{\textregistered}}

\author{Cristian G. Arsene}
\email{christian.arsene@ptb.de}

\author{Dirk Schulze}
\affiliation[Physikalisch-Technische Bundesanstalt, D-38116 Braunschweig, Germany]{Physikalisch-Technische Bundesanstalt (PTB)}

\author{J\"urgen Kratzsch}
\affiliation[Institute for Laboratory Medicine, Clinical Chemistry and Molecular Diagnostics, University of Leipzig, D-04103 Leipzig, Germany]{Institute for Laboratory Medicine, Clinical~Chemistry and Molecular Diagnostics, University of Leipzig}

\author{Andr\'e Henrion}
\affiliation[Physikalisch-Technische Bundesanstalt, D-38116 Braunschweig, Germany]{Physikalisch-Technische Bundesanstalt (PTB)}

\title
{High~Sensitivity~Mass~Spectrometric Quantification~of ~Serum~Growth~Hormone 
by~Amphiphilic~Peptide~Conjugation}

\begin{document}
\centerline {Preliminary draft, May 22, 2012}


\begin{abstract}
\noindent Amphiphilic peptide conjugation affords a significant increase in sensitivity with protein quantification by electrospray-ionization mass spectrometry. This has been demonstrated here for human growth hormone in serum using $N$-(3-iodopropyl)-$N,N,N$-dimethyl\-octyl\-ammo\-nium iodide (IPDOA-iodide) as derivatizing reagent. The signal enhancement achieved in comparison to the method without derivatization enables extension of the applicable concentration range down to the very low concentrations as encountered with clinical glucose suppression tests for patients with acromegaly. The method has been validated using a set of serum samples spiked with known amounts of recombinant 22~kDa growth hormone in the range of 0.48 to 7.65 $\mu$g/L. The coefficient of variation (CV) calculated, based on the deviation of results from the expected concentrations, was 3.5\% and the limit of quantification (LoQ) was determined as 0.4~$\mu$g/L. The potential of the method as a tool in clinical practice has been demonstrated with patient samples of about 1~$\mu$g/L.
\end{abstract}

\section{Introduction}
The reliability of data regarding growth hormone (GH) levels in human serum is crucial for diagnosis and decision-making in the treatment of several diseases, most prominently GH deficiency, gigantism and acro\-megaly. Poor comparability between testing laboratories has motivated the development of isotope dilution mass spectrometry (IDMS) as an alternative to immunoassays\cite{Arsene2010}. The method offers an option to the acquisition of GH target values for serum materials as currently used in ring trials\cite{RFB2009} for quality assurance of clinical testing laboratories, where concentrations typically are well above 5~$\mu$g/L.\par

However, for the method to be of practical use with real patient samples, further improvement in sensitivity was required. GH excretion is mainly pulsatile. High levels of the hormone are only seen immediately after the pulse, whereas concentrations are much lower in the nadir and non-secreting state. For a diagnosis of acromegaly, basal GH has to be measured, which should be found in a wide concentration range higher than 1~$\mu$g/L in case of acromegaly, or can be excluded by levels of less than 0.4~$\mu$g/L in healthy individuals \cite{Chanson2008}. In borderline cases, the suppression of GH during oral glucose tolerance tests is used to distinguish between healthy subjects and patients with active acromegaly. Upper cutoff limits have been reported, which vary between about 1-3~$\mu$g/L depending 
on the assay used\cite{Giustina2000,Barkan2004,Markkanen2006}.

IDMS offers an approach to a SI-traceable and antibody-independent redefinition of the decision limit. It is demonstrated here that the improvement of the signal-to-noise ratio needed to further extend the accessible concentration range down to 1~$\mu$g/L, and less, can be gained by a chemical conjugation of the tryptic signature peptide used for quantification. $N$-(3-iodopropyl)-$N,N,N$-dimethyloctylammonium iodide (IPDOA-iodide) has been synthesized for this purpose and used as a derivatizing reagent in the present study. 
\newpage
\section{Experimental Section} 

{\bf Materials.} {\em GH Reference Solution.} The preparation of this in-house reference material has been described in a previous report\cite{Arsene2008}. It is a solution of recombinant 22~kDa GH (Swiss-Prot P01241-1, A26M, aa26-217) in acetonitrile/water (1:1, v/v) which had been obtained from ProSpec-Tany TechnoGene (Rehovot, Israel). GH concentration is 30.8~nmol/g (686~$\mu$g/g).\par

{\em Isotopically Labeled GH.} Growth hormone, 22 kDa isoform, U-\ce{^{15}}N labeled, 98\%, was kindly provided by Jens Breinholt, Novo Nordisk, M\aa l\o v, Denmark. This  will be referred to as GH*. A solution of GH* in acetonitrile/water (1:1, v/v) was prepared as stock to provide the internal standards needed for IDMS-analysis. The concentration of this solution was determined by the comparison of mass spectrometric signals of tryptic GH* fragments to those of GH in-house standard which had been added in a known amount. GH* concentration was found to be 27.3~nmol/g (611~$\mu$g/g) with the stock solution.\par

{\em GH-Fragment T6.} Peptide YSFLQNPQTSLCFSESIPTPSNR, corresponding to Swiss-Prot P01241-1, aa68-90, had been obtained by custom synthesis from Thermo Fisher (Ulm, Germany). A~solution of T6 (107 nmol/g, i.e., 280~$\mu$g/g, in aqua) was prepared from this and used as stock solution for the experiments within this study.\par

{\em GH-Depleted Serum.} The frozen human serum material was from SunnyLab (SCIPAC), Sittingbourne, UK. It is pooled serum. The product code is SF220-2, batch number 172-156. GH level according to the supplier's certificate is less than 0.1~$\mu$g/L. \par

{\em Calibrators.} Calibrators were prepared by spiking the appropriate amount of GH reference solution into 500 $\mu$L of depleted serum. \par

{\em Samples for Method Validation.} Validation samples were gravimetrically prepared from a stock solution of GH at 8.0~$\mu$g/L. The stock solution at this concentration was obtained by adding an aliquot of GH reference solution to 500~$\mu$L of buffer (20~mM~K\ce{_2HPO_4}, 5~mM~EDTA-\ce{Na^+}, 0.5\%~bovine serum albumin, pH 8.0), lyophilizing and reconstituting with the appropriate amount of depleted serum. Validation samples at the working concentrations were obtained by the appropriate dilution of an aliquot of that stock solution with depleted serum.\par

{\em Clinical Samples.} For method demonstration, anonymized serum surplus from routine samples of the Institute of Laboratory Medicine,  Clinical Chemistry and Molecular Diagnostics, with GH levels between 0.5-1.5~$\mu$g/L (as measured  by Immulite\registered~2000 assay), was used.
The ethics committee of the Leipzig University had no objections to the study (082-10-19042010).
Samples were stored at -80~$^\circ$C and conditioned for at least 1~h at room temperature prior to use.\par

{\em $N$-(3-iodopropyl)-$N,N,N$-dimethyloctylammonium Iodide,  \ce{[(IC_3H_6)N^+(CH_3)_2(C_8H_{17})]I^-}}.\newline 
Alkylating reagent, IPDOA-iodide, was prepared by the dropwise addition of a solution of $N,N$-di\-methyl\-octyl\-amine\  (0.782~g, 5~mmol, in 10~mL $tert$-butylmethyl ether) to 1,3-diiodopropane (2.959~g, 10~mmol). The diiodo\-propane was kept at 50~$^\circ$C during the addition of the amine, which was brought to completion within 3~h. The mixture was then stirred at room temperature for another 48~h. The yellow phase which had formed was separated from the supernatant and washed with $tert$-butylmethyl~ether. For further purification, the product was first collected with $tert$-butylmethyl~ether to which as much methanol as needed for dissolution was added, and then re-precipitated by the addition of $tert$-butyl\-methyl~ether. The substance could not be gained as crystals but formed a viscous oil. The reaction yield was 1.687~g, 3.7~mmol (74\%). The identity was verified by the assignment of $^1$H- and $^{13}$C-NMR signals and by checking the molecular mass of the ammonium ion: $^1$H-NMR (\ce{CDCl_3}, 400~MHz): $\delta$, ppm: 0.88 (t, $J=$ 6.6~Hz, 3H, CH\ce{_3}), 1.22-1.34 (m, 6H, CH\ce {_2}), 1.34-1.46 (m, 4H, CH\ce{_2}), 1.75-1.83 (m, 2H, CH\ce{_2}), 2.34-2.42 (m, 2H, CH\ce{_2}), 3.32 (t, $J=$ 6.4~Hz, 2H, CH\ce{_2}), 3.42-3.47 (m, 6H, CH\ce{_3}), 3.50-3.54 (m, 2H, CH\ce{_2}), 3.73-3.77 (m, 2H, CH\ce{_2}). Similarly, $^{13}$C-NMR signals are in agreement with the assumed structure. The monoisotopic mass was determined by time-of-flight mass spectrometry as 326.1358~Da  for \ce{(IC_3H_6)N^+(CH_3)_2(C_8H_{17})}, which is 6~ppm off the calculated mass (326.1339~Da).\par

{\bf GH-Quantification Method.} {\em Setting up Sample and Calibrator.} 500 $\mu$L of serum sample were used for each run.  With each sample, a calibrator was prepared at a GH concentration, matching as closely as possible the concentration expected in that sample. Then both sample and calibrator were spiked with equal amounts of GH* (isotopically labeled GH). The amount of GH* also was chosen so as to  match the concentration of GH in both sample and calibrator.\par 

{\em Sample Pretreatment.} After spike-equilibration, sample and calibrator were subjected to tryptic proteolysis as previously described\cite{Arsene2010}. DTT and water/TFA were added at the end as described, but the reaction of the mixture with iodo\-acet\-amide was left out.\par

{\em Extraction of T6 (T6*) by RP- and SCX- Liquid Chromatography.} Fragments T6 (as formed from GH) and T6* (from GH*) were isolated by reversed phase (RP) chromatography followed by strong-cation exchange (SCX-) chromatography using a Jupiter-RP C18 column (10$\times$250~mm, 300~\AA) and a Luna-SCX column (10$\times$250~mm, 100~\AA, 5~$\mu$m) Phenomenex, Torrance, CA, USA. Solvents used were A: water (0.1\% TFA), B: acetonitrile (0.1\% TFA), C: KH\ce{_2}PO\ce{_4}, 5~mM in acetonitrile/water (1:4, v/v), pH 2.83, D: same as C, but additionally KCl (0.5 M). Solvent gradients were A/B (1\% B at 0-5~min, 1-80\% B within 40~min, 80\% B for 5 min) in RP-chromatography and C/D (0\% D at 0-3~min, 0-100\% D within 37~min, 100\% D for 5 min) in SCX-separation. The flow rate was 1.5~mL/min in both cases. T6 (T6*) containing fractions were automatically collected within 30.35-31.45~min (RP) and 33.2-35.0~min (SCX). Typically, 5 consecutive RP-chromatography runs were required to clean up the whole of the material (850~$\mu$L) resulting from the hydrolysis step. 170~$\mu$L  were injected with each run. Collected fractions were unified, lyophilized, re-dissolved in 300~$\mu$L of TCEP-HCl (10~mM in acetonitrile/water, 1:4, v/v) and kept at 37~$^\circ$C for 1~h prior to injecting the whole amount into the SCX column. The T6(T6*)-containing fraction obtained after SCX-chromatography was desalted using C18-SPE cartridges and lyophilized.\par

{\em Conjugation with IPDOA-Iodide.} The lyophilized product was collected with 50~$\mu$L of TRIS buffer (50~mM TRIS/TRIS-HCL in acetonitrile/water 65:35, pH 8.5, 3~mM DTT). Following incubation for 1~h at 37~$^\circ$C, 4~mg of IPDOA-iodide were added and the mixture kept at room temperature for 24~h. Then, 4~mg cysteine (dissolved in 50~$\mu$L water) and 400~$\mu$L of acetonitrile/water 15:85 (v/v), 0.1\% TFA were added. Eventually, IPDOA-conjugated T6 (T6*) was isolated by another RP-chromatographic clean-up using a Hibar Lichrosphere\registered 300~\AA ~C18 column (10$\times$250~mm, Merck, Darmstadt, Germany). The gradient (solvents A/B, see preceding subsection) was 20\% B at 0-5~min, 20-90\% B within 40 min, 90\% B for 5~min. IPDOA-T6 (T6*) was collected in between 25.5-27.5 ~min. The solution was lyophilized and dissolved in 30~$\mu$L of water/acetonitrile, 9:1 (v/v), 0.1\% formic acid, prior to analysis by LC/MS-MS.\par

{\em Analysis by LC/MS-MS.} For separation, a Discovery BIO wide pore C18 RP column (2.1$\times$150 mm, 3~$\mu$m, 300~\AA) was used. A gradient mixture of solvents A/B (A: water, 0.1\% formic acid, B: acetonitrile, 0.1\% formic acid) was applied for elution with 10\%~B at 0-3~min, 10\%~B to 80\%~B within 37~min, 80\%~B for 5~min. The flow-rate was 0.2~mL/min. The retention time was 23.9~min for IPDOA- modified T6 (T6*). The mass spectrometer (4000 Q Trap, Applied Biosystems, Foster City, CA, USA) was run in unit resolution, and transitions {\em m/z} 939.0$\rightarrow$1031.5 (949.0$\rightarrow$1042.5) were monitored for T6 (T6*). Parameters were optimized for maximum signal intensities. The ESI source was at 4.5~kV sprayer voltage and 350~$^\circ$C dry gas (\ce{N_2}). The collision cell was set to high pressure, collision energy to 42 eV.\par

{\bf Monitoring Time Course of Product Formation IPDOA-T6.} Aliquots of an aqueous T6-solution were used in this experiment, each one containing 23~fmol of T6. These were lyophilized, collected with TRIS- buffer, reacted with IPDOA-iodide, and isolated as given in subsection {\em Conjugation with IPDOA- Iodide}. Reaction times were varied with the aliquots according to the time points shown in Figure  \ref{timecourse} (0, 5, 10, 20, 40, 80, 120, 240, 480, 960 and 1440~min). Two aliquots were run for each time point. Relative amounts of IPDOA-T6 as plotted in the Figure were obtained by LC/MS-MS using the conditions described in subsection {\em Analysis by LC/MS-MS}, except that only transition {\em m/z} 939.0$\rightarrow$1031.5 was recorded, corresponding to only T6 (not T6*).

{\bf Characterization of IPDOA-Conjugated T6 by Mass Spectrometry.} IPDOA-conjugated and carbamidomethylated T6, respectively, which had been prepared at 20~pmol/g in acetonitrile/water (1:1, v/v), 0.1 M acetic acid, were injected at 5~$\mu$L/min into the electrospray-source of a micrOTOF-Q instrument (Bruker Daltonik GmbH, Bremen, Germany). The source was at 4.5~kV capillary voltage and 250 $^\circ$C dry gas  temperature. Collision energy was set to 35~eV (Ar) for the fragmentation experiment, otherwise to 10~eV.\par 

{\bf Comparing the Matrix Effect on LC/MS-MS signals of IPDOA-T6 and CAM-T6.} The experiment is as described in section {\em Results}, procedures used are as given in subsection {\em GH-Quan\-ti\-fi\-ca\-tion Method}. However, with analysis by LC/MS-MS, in place of transition {\em m/z} 949.0$\rightarrow$1042.5 (for T6*),  {\em m/z} 892.3$\rightarrow$ 671.3 was monitored here to allow for CAM-T6 to be observed, which was eluting at  20.9 min.

\section{Results}

\begin{figure}
\includegraphics{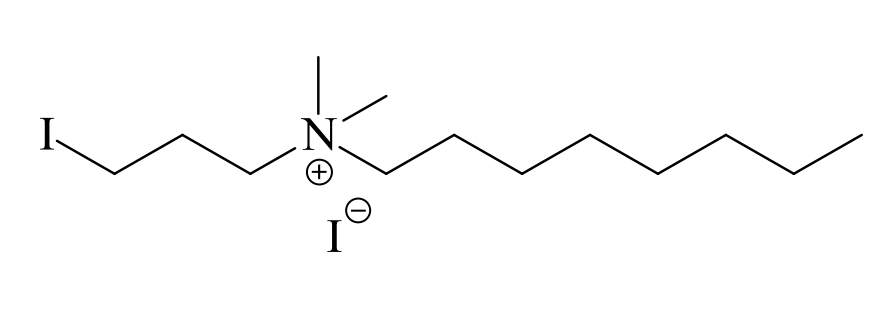}
\caption{$N$-(3-iodopropyl)-$N,N,N$-dimethyloctylammonium iodide.} \label{IPDOA}
\end{figure}

\begin{figure}[t]
\includegraphics{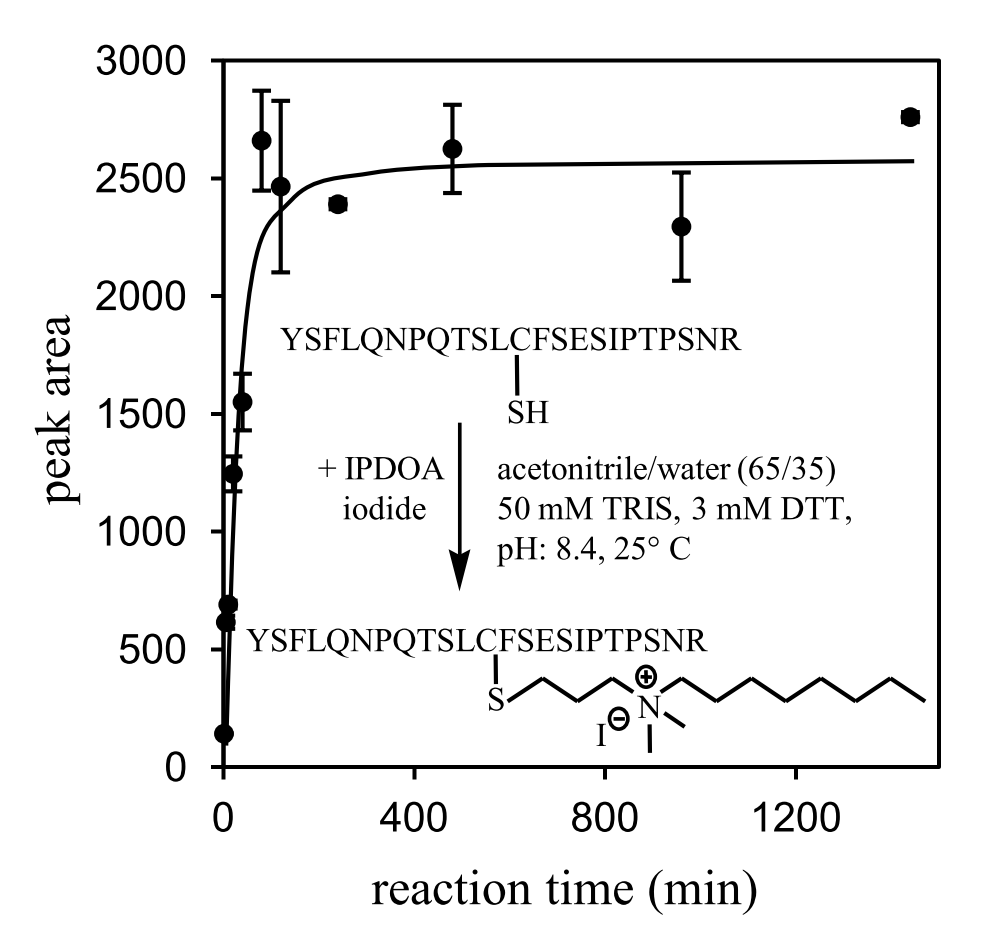}
\caption{Time course of T6-peptide conjugation by IPDOA-iodide. LC/MS-MS peak areas for transition ({\em m/z} 939.0$\rightarrow$1031.5) were used to represent the relative amounts of IPDOA-T6. Plotted data are averages of two runs, and error bars are the associated standard deviations. }\label{timecourse}
\end{figure}

{\bf IPDOA-Iodide.} IPDOA-iodide, Figure \ref{IPDOA}, was obtained by reacting the respective amine base with 1,3-diiodopropane as described by Hofmann\cite{Hofmann1860} and Litterscheid\cite{Litterscheid1904}. This reagent, similar to iodo\-acet\-amide, modifies thiol-containing peptides in simple reaction at room temperature and within less than 1-2~h. This is exemplified for the tryptic fragment T6 of GH in Figure \ref{timecourse}.

\begin{figure}[]
\includegraphics{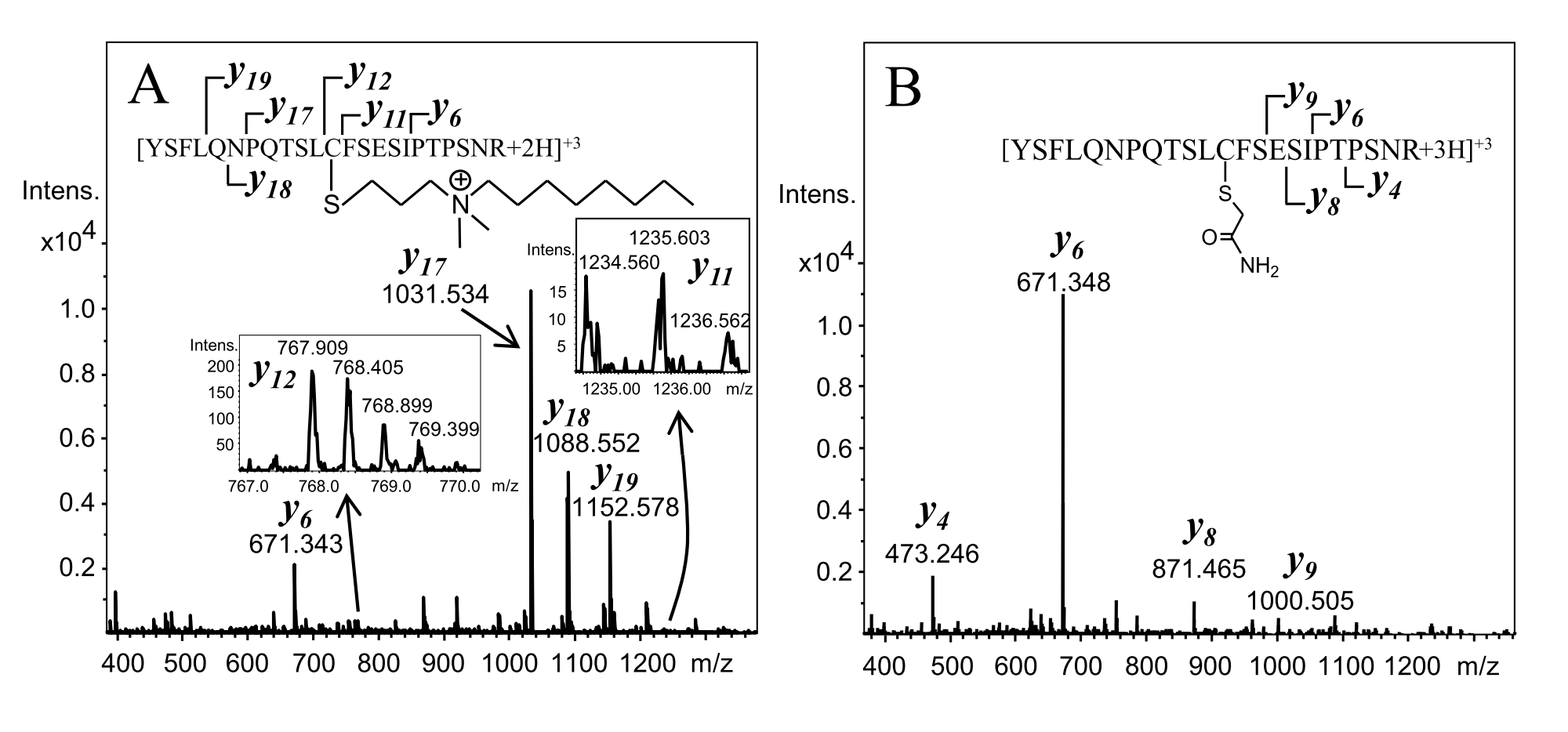}
\caption{CID-Fragmentation spectra of T6 (A) IPDOA-conjugated $[($IPD\-OA-T6$)$+2H$]^{3+}$, and (B) carbamidomethylated $[($CAM-T6$)$+3H$]^{3+}$.}\label{CIDspectrum}
\end{figure}

The most abundant ion observed for IPDOA-conjugated T6  was $[($IPD\-OA-T6$)$+2H$]^{3+}$ under electrospray conditions which had been optimized for signal abundance. The CID-frag\-men\-tation spectrum of this ion is shown in Figure \ref{CIDspectrum}A.

By the masses of \ce{$y_{11}$} (the last fragment in the series not yet containing cysteine) and the first fragment that includes cysteine (\ce{$y_{12}$}), even though the signals are of low abundance, it is evident that IPDOA has conjugated T6 at cysteine. Also, ions in the {\em y-}series that correspond to cysteine-containing fragment ions (namely \ce{$y_{17}$}, \ce{$y_{18}$}, \ce{$y_{19}$}) are doubly charged, while the shorter ones (most prominently \ce{$y_{6}$}) are singly charged. This is in striking contrast to the behavior of carbamidomethylated T6 (CAM-T6, Figure \ref{CIDspectrum}B), which produces exclusively singly charged fragment ions. Obviously, a permanent positive charge is located at the IPDOA-conjugated cysteine, as expected.\par

\begin{figure}[]
\includegraphics{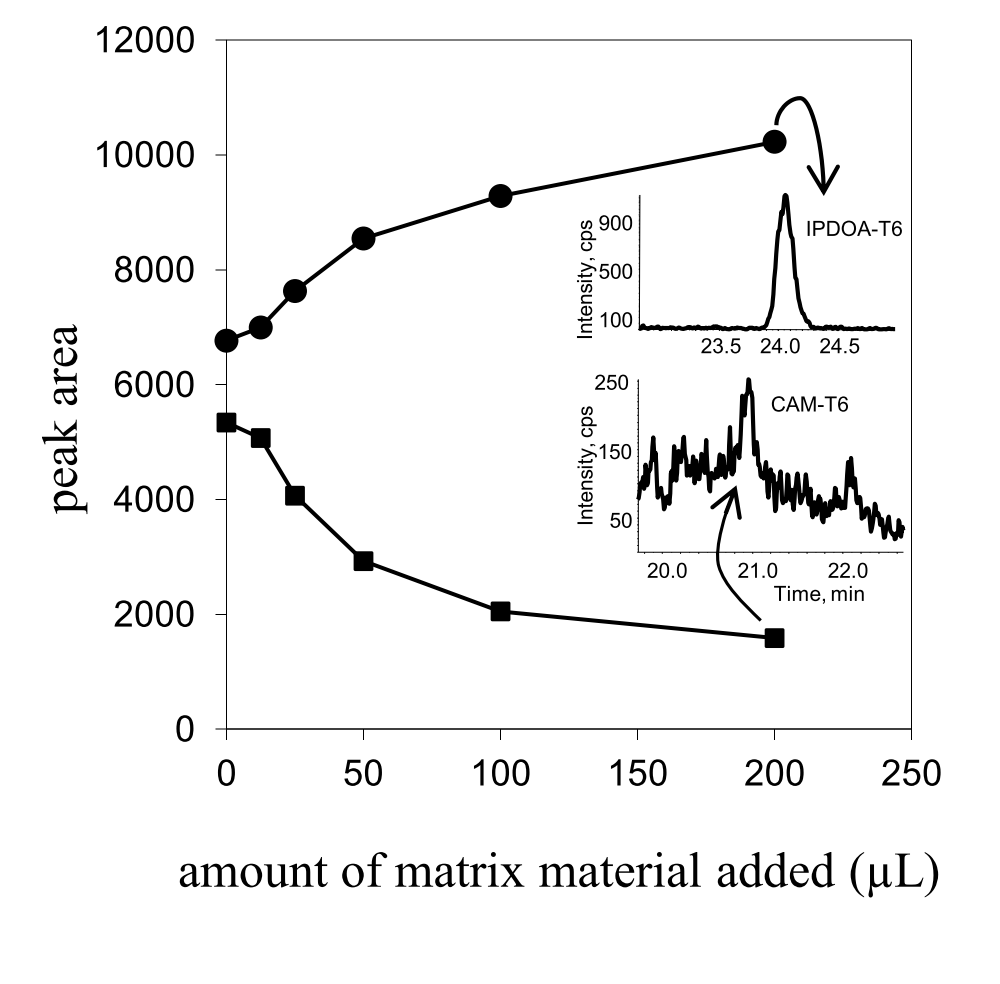}
\caption{Effect of matrix on ESI-LC/MS-MS signal abundance with IPDOA-T6 (circles) and CAM-T6 (squares). At maximum addition (200~$\mu$L) the amount of matrix corresponds to what is present after the clean-up by RP-LC and SCX-chromatography in a typical analysis. Insets: Signals obtained at this point.}\label{matrixsuppression}
\end{figure}

{\bf Matrix Effect on Signal Intensities of Carbamidomethylated vs. IPDOA-Conjugated T6-Peptide.} This experiment was conducted to characterize the influence of matrix-load on the mass spectrometric responses of both IPDOA-T6 and, for comparison, on carbamidomethylated T6 (CAM-T6). Conditions were chosen to be representative for the composition of a sample extract after chromatographic clean-up according to the GH-quantification method given in this paper. To this end, GH-depleted serum (500~$\mu$L) was subjected to the extraction procedure by RP-LC and subsequent SCX-chromatography as described in the {\em Experimental Section}. In particular, the fractions that normally would contain T6 (T6*) were collected (T6* is referring to the isotopically labeled T6). The "cleaned-up" serum material obtained this way would be used to mimic remaining matrix impurities still present at this stage of the analysis procedure in spite of the clean-up. In Figure \ref{matrixsuppression} the effect of this material on LC/MS-MS signals for IPDOA-T6 and CAM-T6 is shown. Starting with matrix-free solutions of both conjugates, increasing amounts of the matrix material were added while the total content of T6 (IPDOA- and CAM-) was kept constant at 23~fmol with all samples. In this way, T6-levels were such as would be obtained from a serum sample at about 1~$\mu$g/L. While in matrix-free solution the signal acquired for IPDOA-T6 is not much different from what is seen with CAM-T6, there is about a factor 6 between signal intensities in presence of a realistic level of matrix-background (200~$\mu$L).\par

\newpage

\begin{figure}[t]
\includegraphics{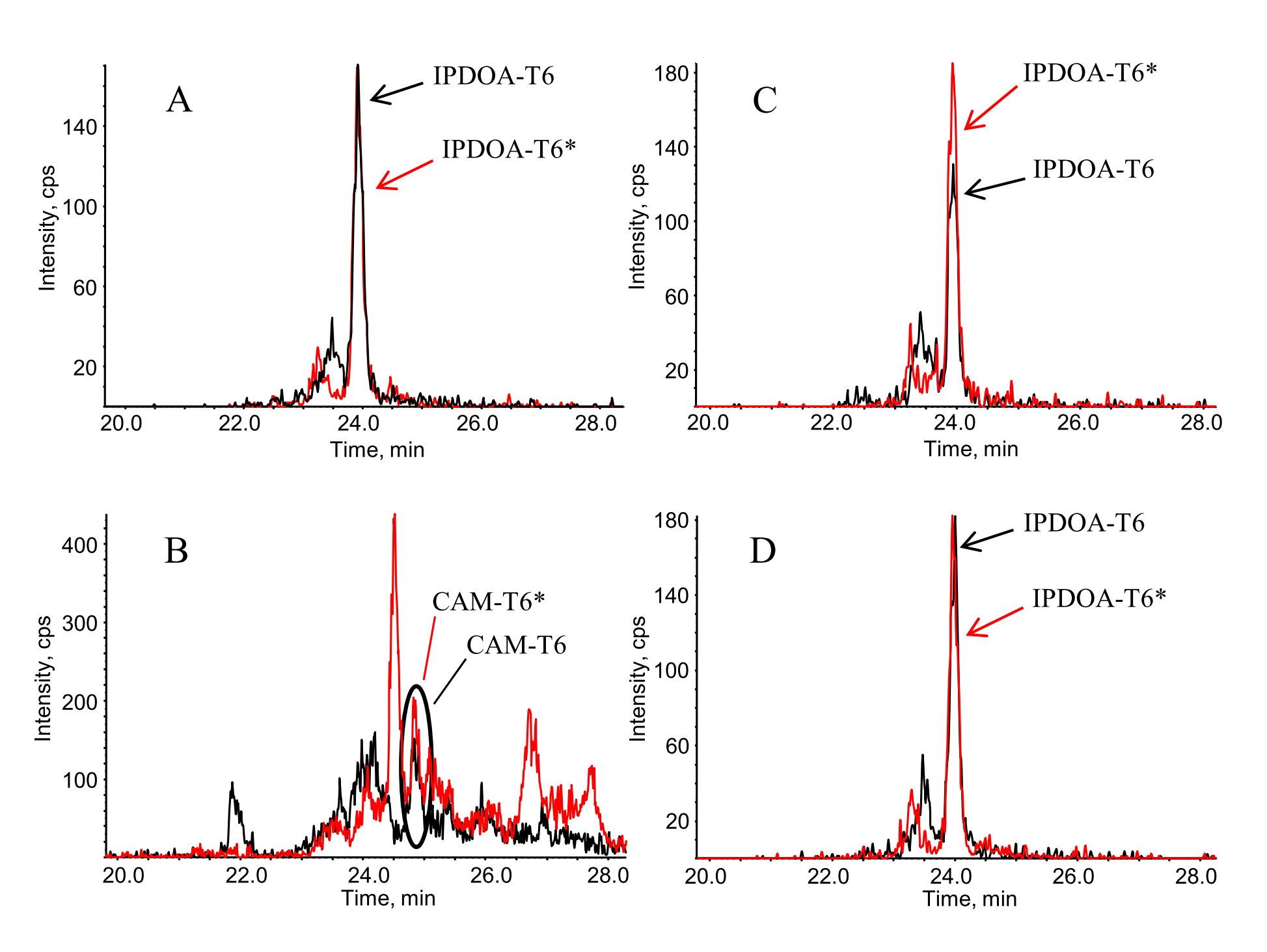}
\caption{Instrumental responses obtained for a blank serum spiked with GH at 0.95~$\mu$g/L using (A) the IPDOA-conjugation method given in this report and (B) the CAM-conjugation method previously described. For comparison, the results obtained for patient sera at a similar concentration are shown, (C) and (D). MS-MS transitions recorded were {\em m/z} 939.0$\rightarrow$1031.5 (949.0$\rightarrow$1042.5) for IPDOA-T6 (IPDOA-T6*) and {\em m/z} 892.3$\rightarrow$ 671.3 (902.3$\rightarrow$681.3) for CAM-T6 (CAM-T6*). }\label{comparisonofresponses}
\end{figure}

{\bf GH-Quantification Method.} The entire method can be summerized as follows:
\begin{enumerate}
\item{Spike the serum sample (500~$\mu$L) with U-\ce{^{15}}N labeled GH,  (GH*). The same with the respective calibrator (GH-depleted serum plus a known amount of GH). With both:}
\item{Proteolyze (trypsin, 37 $^\circ$C). Lyophilize. Collect with water (2\% trifluoroacetic acid).}
\item{Extract T6, T6* by (1) C18- Reversed phase liquid chromatography (RP-LC) followed by (2) strong cation exchange chromatography (SCX).}
\item{React with IPDOA-iodide. Remove excess of IPDOA-iodide from the product by means of another RP-LC clean-up.}
\item{Acquire IPDOA-T6/ IPDOA-T6* ratios by analytical RP-LC/MS-MS.}
\item{Calculate GH amount in sample from signal ratio obtained for the serum sample relative to the ratio obtained for the calibrator, taking into account the known amount of GH in the calibrator.}
\end{enumerate}
A series of samples (depleted serum spiked with known amounts of 22 kDa GH) was used for the validation of this method. The results are compiled in Table \ref{recoverytable}. As an example for illustration, the LC/MS-MS signal obtained for the 0.95~$\mu$g/L sample  is shown in Figure \ref{comparisonofresponses}A, the corresponding result with the T6-CAM method, for comparison, in Figure  \ref{comparisonofresponses}B. In order to check if the method is producing useful results with real samples too, a set of native sera from healthy donors at basis- levels of GH concentration were analysed in duplicate. Results ($\mu$g/L) were: 0.69$\pm$0.04; 0.82$\pm$0.03; 1.50$\pm$0.05; 1.00$\pm$0.06. LC/MS-MS responses for two of these sera are shown in Figure \ref{comparisonofresponses}, (C) and (D).

\rule{0pt}{30pt}
\begin{table}[t]
\begin{tabular}{l} \\
\begin{tabular}{ccr}
expected, & found, & recovery,\\ 
 $\mu$g/L  &$\mu$g/L &  \%  \\ 
&&\\
0.48       & 0.53   &   109.3   \\
0.48       & 0.51   &   105.9   \\
0.95       & 0.94   &    99.0    \\
0.95       & 0.96   &   100.9   \\
1.90       & 1.88   &    98.7    \\
1.90       & 1.92   &   101.1   \\
3.83       & 3.75   &   98.0     \\
3.83       & 3.80   &   102.9   \\
7.65       & 7.71   &   100.7   \\
7.65       & 7.75   &   102.5   \\
&&\\ 
\end{tabular}
\\
\begin{tabular}{lr}
 
mean recovery, \%   & 101.4\\
CV, \%                    & 3.5    \\
LoQ$^a$, $\mu$g/L        & 0.4    \\
R$^2 $(linear regression)                    & 0.99   \\  
  &\\ \hline
\multicolumn{2}{l}{$^a$  calculated from linear} \\
\multicolumn{2}{l}{\rule{8pt}{0pt}regression data} \\                  
\end{tabular}
\end{tabular}

\rule{0pt}{30pt}\caption{Recovery of defined additions of GH to depleted serum.} \label{recoverytable}
\end{table}

\newpage

\section{Discussion}

\begin{figure}[]
\includegraphics{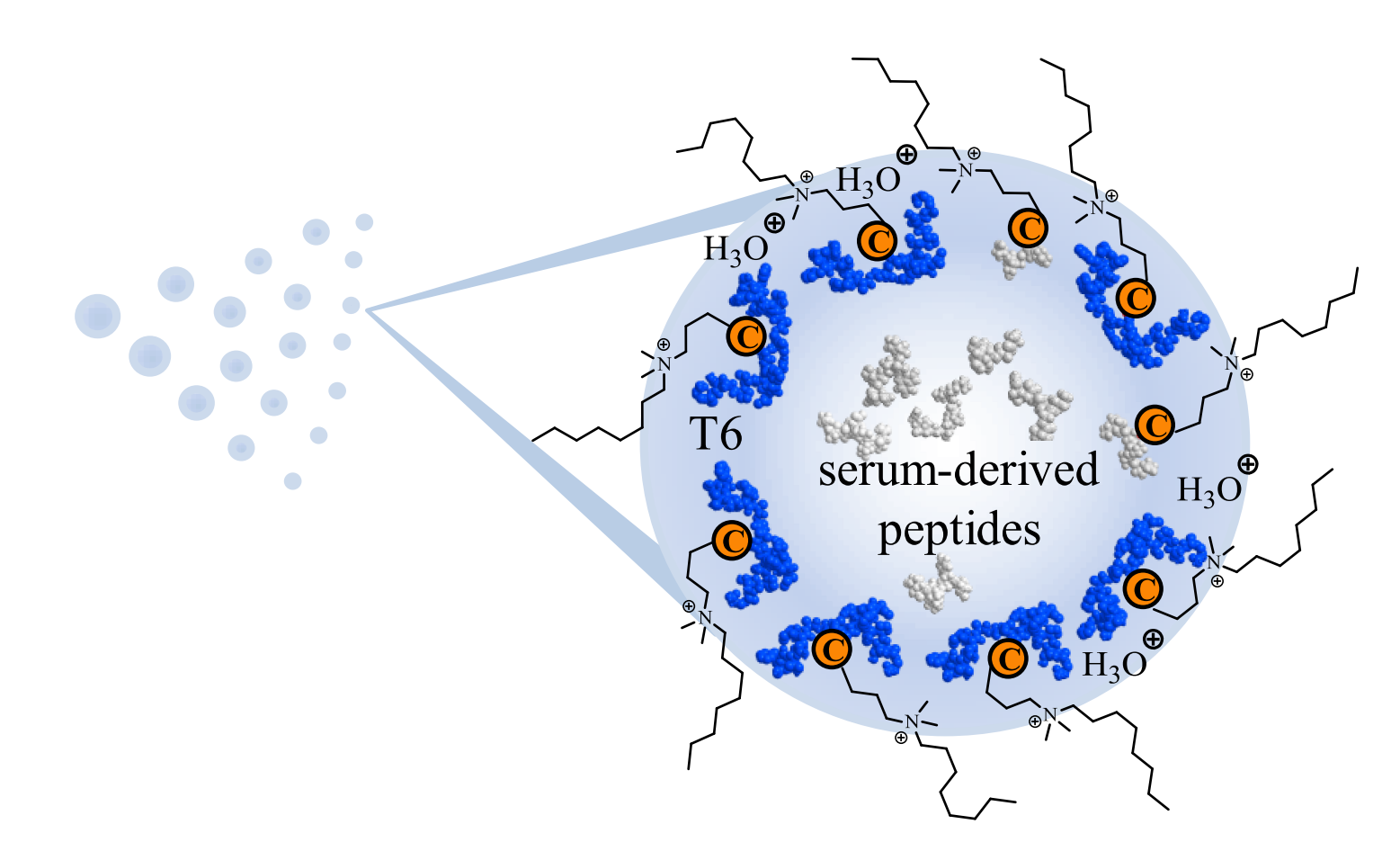}
\caption{(Scheme) Enrichment of T6 at the charge-carrying ESI-droplet surface as a result of IPDOA-conjugation. Serum-derived peptides, as long as not cysteine-containing are not reacting and assumed to be discriminated therefore in competition about the charge.}\label{esidroplet}
\end{figure}

{\bf Conjugation Reagent.} Chemical conjugation to increasing the abundance of targeted analyte ions in mass spectrometry has been a concept of long standing. For instance, Chai~et~al.\cite{Chai1987} in 1987 have reported by a factor of 4-29 better sensitivities in the detection of amino acids by liquid-assisted fast atom bombardment (FAB) on di-isopropylphosphorylation of N-terminal and side-chain amino groups. Two main features to be imposed on the substrate by derivatization had been re\-cog\-nized  as promoting its ionization relative to non derivatized matrix components: a pre-existing permanent charge on the molecule\cite{Busch1982} and increased surface activity\cite{Ligon1986}. Mass spectrometric responses with electrospray ionization (ESI) have proven to be subject to the same rules. Most of the results published in the area are based either on increasing hydrophobicity (and in this way, surface activity of the substrate)\cite{Ullmer2006, Foettinger2006, Frahm2007, Williams2008, Williams2009, Shuford2010, Kulevich2010}, increasing basicity so as to promote protonation in solution\cite{Shortreed2006}, or on introducing a permanent positive charge in addition to non-polar moieties\cite{Yang2006, Mirzaei2006}. IPDOA-iodide (Figure \ref{IPDOA}) has been designed as a new reagent featuring a quaternary ammonium site next to a longer aliphatic moiety, which lends the molecule both of the properties believed to be necessary for optimal promotion of ion formation. Unlike previous work using NHS-esters of carboxylic acids as reagents, IPDOA-iodide only reacts with peptides containing cysteine. These are then selectively enhanced, while excluding the rest of the peptides (Figure \ref{esidroplet}). This was expected to be an advantage over reagents uniformly modifying all peptides, particularly so if working with complex matrices as serum. 
On the other hand, of course, the set of tryptic signature peptides available for quantification of an analyte protein is correspondingly confined in this way.\par

{\bf Effect of IPDOA-Conjugation.} Gains in signal abundance have been reported in ranges from 2-4 fold\cite{Frahm2007, Williams2009}, or 1.2-34 fold\cite{Kulevich2010}, up to even more than a factor of 2000\cite{Williams2008} with solutions of model peptides. The improvement obtained here by using IPDOA-T6 is just a factor of 5-6. However, in contrast to previous work, this is a result from an experiment in presence of a background of peptides, as is to be expected realistically with serum samples in quantitative analysis by LC/MS-MS. Interestingly, signals increase with the addition of matrix-peptides to IPDOA-T6 as opposed to CAM-T6, where the reverse effect is observed (Figure \ref{matrixsuppression}). Referring to the equilibrium partitioning model by Cech \& Enke\cite{Cech2000, Cech2001}, this may lead to the conclusion that IPDOA-conjugated T6 on average is outcompeting the background of serum derived peptides for the excess charge on the ESI-droplet, whereas it is happening the other way round with CAM-T6. The advantage of using IPDOA-iodide is illustrated in Figure \ref{comparisonofresponses}. For a serum sample at a GH concentration of 0.95~$\mu$g/L, which is less than the limit of quantification determined for the CAM-based method (1.7~$\mu$g/L), a signal of well-acceptable signal-to-noise ratio is still obtained for IPDOA-T6 (A), while the signal for CAM-T6 (B) can scarcely be used for quantification.

{\bf Absolute Quantification of Growth Hormone by Isotope Dilution Mass Spectrometry.} Strategies for extracting cleavage fragments of interest from body fluids for subsequent analysis by IDMS have been reviewed recently by Halquist \& Karnes\cite{Halquist2011}. In principle, it appears reasonable and easier to first isolate or enrich as much as possible the targeted protein itself (plus isotopically labeled protein used as internal standard) in order to reduce complexity prior to tryptic proteolysis and LC/MS-MS. In this study, however, the alternative of first completely proteolyzing the serum sample and then extracting the signature peptides from the mixture, proved simple and reproducible according to the developed protocol. In a previous investigation (development of the CAM-T6 based method)\cite{Arsene2010} it was found that the extraction of the peptide from the proteolysate by reversed phase chromatography (RP-LC) followed by further clean-up using strong cation-exchange (SCX-) chromatography would yield better results  than RP-LC alone, in terms of signal-to-noise ratio. This was not changed in designing the present method. However, deviating from the original method, where conjugation by iodoacetamide is done directly after proteolysis, with the IPDOA-T6 protocol, the (reduced, but non derivatized) T6 (T6*) fraction is first isolated from the proteolysate and only then IPDOA-conjugated. This is owing to the fact that, IPDOA-iodide, otherwise present in excess as an ion pairing agent, would be severely compromising separation by RP-LC as well as by SCX-chromatography. For the same reason, after IPDOA-conjugation of T6(T6*) an additional RP-LC step was introduced in order to cut away excess of reagent prior to analytical LC/MS-MS.\par

The data compiled in Table \ref{recoverytable} indicate the reliability of the analytical method down to GH serum concentrations of about 0.4~$\mu$g/L, which was determined as being the limit of quantification (LoQ). The improvement achieved by IPDOA-conjugation enables the precise determination of GH in the range relevant with patient samples collected in glucose tolerance tests, as was the aim of the study. Responses obtained for the native samples, see examples shown in Figure \ref{comparisonofresponses}, (C) and (D), are of the same quality as those acquired for the validation samples (see Figure \ref{comparisonofresponses} A, C and D). This suggests the robustness of the method against varying background interferences resulting from slight patient-to-patient variations in serum (protein-) composition as expected in clinical practice.\par
Unlike previous feasibility studies, which had demonstrated the detectability of plasma GH either at concentration levels far above physiological ($\geq 1000~\mu$g/L)\cite{Yang2007} or with model samples of unrealistically diluted plasma background \cite{Wu2002}, the method described here is applicable to native patient samples for diagnostic purposes.

\section{Conclusion}
Absolute quantification by LC/ESI-IDMS, owing to a reliability inherent to the measurement principle, is generally attributed to the potential of providing the basis for the development of reference methods in quantification of protein health status markers, if not fully replacing antibody-based methods. 
Particularly for growth hormone related disorders, the clinical decision-making is not urgent. This enables the use of mass spectrometry for the most accurate determination of growth hormone for clinical applications as e.g. oral glucose tolerance tests. It has been demonstrated here, using the example of growth hormone measurement, that selective conjugation of the analyte makes a contribution to improving the sensitivity of mass spectrometric determination as much as rendering it competitive to immunoassays. The reagent used for this, IPDOA-iodide, is by design not just enhancing the ionization yield of the peptide employed for GH quantification (T6), but also selectively modifying only cysteine containing peptides. This strongly reduces the fraction of matrix-derived-peptides being conjugated at the same time and, therefore, competes with IPDOA-T6 in the electrospray process. The work reported here is constituting the first application of peptide conjugation affording improved limits of detection for mass spectrometric quantification of proteins from native sera.\par

\acknowledgement
Labeled growth hormone was kindly provided by Jens~Breinholt (Novo Nordisk, M\aa l\o v, Denmark). NMR- measurements were by courtesy of the laboratory of Prof.~Stefan~Schulz (Technische Universit\"at Braunschweig).

\newpage    

\bibliography{amphiconjug}

\end{document}